\documentclass[12pt,notitlepage]{article}
\usepackage{amsfonts, amssymb, amsmath, graphicx}
\usepackage[top=3cm, bottom=3cm, left=3.08cm, right=3.08cm]{geometry}
\usepackage{algorithm, algpseudocode}

\input{common.def}

\begin{document}

\title{Machine Learning Techniques\\for Intrusion Detection}
\author{Mahdi Zamani and Mahnush Movahedi}
\date{\vspace*{-5pt}\small \{zamani,movahedi\}@cs.unm.edu \\ Department of Computer Science \\ University of New Mexico}
\maketitle

\begin{abstract}
An Intrusion Detection System (IDS) is a software that monitors a single or a network of computers for malicious activities (attacks) that are aimed at stealing or censoring information or corrupting network protocols. Most techniques used in today's IDS are not able to deal with the dynamic and complex nature of cyber attacks on computer networks. Hence, efficient adaptive methods like various techniques of machine learning can result in higher detection rates, lower false alarm rates and reasonable computation and communication costs. In this paper, we study several such schemes and compare their performance. We divide the schemes into methods based on classical artificial intelligence (AI) and methods based on computational intelligence (CI). We explain how various characteristics of CI techniques can be used to build efficient IDS.
\end{abstract}

\fontsize{12pt}{15pt}\selectfont

\section{Introduction}
Today, political and commercial entities are increasingly engaging in sophisticated cyber-warfare to damage, disrupt, or censor information content in computer networks~\cite{cyberwarfare2012}. In designing network protocols, there is a need to ensure reliability against intrusions of powerful attackers that can even control a fraction of parties in the network. The controlled parties can launch both passive (\eg eavesdropping, non-participation) and active attacks (\eg jamming, message dropping, corruption, and forging).

Intrusion detection is the process of dynamically monitoring events occurring in a computer system or network, analyzing them for signs of possible incidents and often interdicting the unauthorized access~\cite{scarfone:nist:2007}. This is typically accomplished by automatically collecting information from a variety of systems and network sources, and then analyzing the information for possible security problems. 

\parsec{Motivation} Traditional intrusion detection and prevention techniques, like firewalls, access control mechanisms, and encryptions, have several limitations in fully protecting networks and systems from increasingly sophisticated attacks like denial of service. Moreover, most systems built based on such techniques suffer from high false positive and false negative detection rates and the lack of continuously adapting to changing malicious behaviors. In the past decade, however, several Machine Learning (ML) techniques have been applied to the problem of intrusion detection with the hope of improving detection rates and adaptability. These techniques are often used to keep the attack knowledge bases up-to-date and comprehensive.

\parsec{Study Approach} In this paper, we study several papers that use ML methods for detecting malicious behavior in distributed computer systems. There is a huge body of work in this area thus, we decided to carefully select a few papers based on two factors: diversity and citations count. By diversity we mean most ML techniques for IDS are covered but only one paper is picked from the set of papers that use the same technique. Also, the papers are chosen based on their citations count as this factor greatly shows how much the corresponding work has influenced the community. All non-survey papers studied here are cited at least 100 times. 


\parsec{Paper Organization} In Section~\ref{sec:ID-Challenges}, we briefly state the main challenges in intrusion detection and describe two general approaches for solving these problems. In Section~\ref{sec:AI-Techniques}, we review several intrusion detection techniques based on traditional AI. In section~\ref{sec:CI-Techniques}, we define various core methods of computational intelligence and describe several CI-based algorithms proposed in the literature.  

\section{Challenges and Approaches}
\label{sec:ID-Challenges}
An IDS generally has to deal with problems such as large network traffic volumes, highly uneven data distribution, the difficulty to realize decision boundaries between normal and abnormal behavior, and a requirement for continuous adaptation to a constantly changing environment~\cite{Tsai:2009:RID:1598090.1598543}. In general, the challenge is to efficiently capture and classify various behaviors in a computer network.
Strategies for classification of network behaviors are typically divided into two categories: \emph{misuse detection} and \emph{anomaly detection}~\cite{scarfone:nist:2007}.

Misuse detection techniques examine both network and system activity for known instances of misuse using signature matching algorithms. This technique is effective at detecting attacks that are already known. However, novel attacks are often missed giving rise to false negatives. Alerts may be generated by the IDS, but reaction to every alert wastes time and resources leading to instability of the system. To overcome this problem, IDS should not start elimination procedure as soon as the first symptom has been detected but rather it should be patient enough to collect alerts and decide based on the correlation of them.

Anomaly detection systems rely on constructing a model of user behavior that is considered normal. This is achieved by using a combination of statistical or machine learning methods to examine network traffic or system calls and processes. The detection of novel attacks is more successful using the anomaly detection approach as any deviant behavior is classified as an intrusion. However, normal behavior in a large and dynamic system is not well defined and it changes over the time. This often results in a substantial number of false alarms known as false positives. A network-based IDS looks at the incoming network traffic for patterns that can signify whether a person is probing the network for vulnerable computers. Since responding to each alert consumes relatively large amounts of time and resources, IDS should not respond to every alert it generates. Disregarding this fact may result in a self-inflicted denial-of-service. To overcome this problem, alerts should be aggregated and correlated in order to produce fewer but more expressive and remarkable alerts.

\parsec{Machine Learning Approaches} We divide the ML-based approaches to intrusion detection into two categories: approaches based on \emph{Artificial Intelligence (AI)} techniques and approaches based on \emph{Computational Intelligence (CI)} methods. AI techniques refer to the methods from the domain of classical AI like statistical modeling and while CI techniques refer to nature-inspired methods that are used to deal with complex problems that classical methods are unable to solve. Important CI methodologies are \emph{evolutionary computation}, \emph{fuzzy logic}, \emph{artificial neural networks}, and \emph{artificial immune systems}. CI is different from the well-known field of AI. AI handles symbolic knowledge representation, while CI handles numeric representation of information. Although the boundary between these two categories is not always clear and many hybrid methods have been proposed in the literature, most previous work are mainly designed based on either of the categories. Moreover, it would be quite useful to understand how well nature-based techniques perform in contrast to classical methods. 

\section{AI-based Techniques} 
\label{sec:AI-Techniques}
Laskov \etal~\cite{Laskov:2005} develop an experimental framework for comparative analysis of supervised (classification) and unsupervised learning (clustering) techniques for detecting malicious activities. The supervised methods evaluated in this work include \emph{decision trees}, \emph{k-Nearest Neighbor (kNN)}, \emph{Multi-Layer Perceptron (MLP)}, and \emph{Support Vector Machines (SVM)}. The unsupervised algorithms include \emph{$\gamma$-algorithm}, \emph{k-means clustering}, and \emph{single linkage clustering}.
They define two scenarios for evaluating the aforementioned learning algorithms from both categories. In the first scenario, they assume that training and test data come from the same unknown distribution. 
In the second scenario, they consider the case where the test data comes from new (\ie unseen) attack patterns. This scenario helps us understand how much an IDS can generalize its knowledge to new malicious patterns, which is often very essential for an IDS system since today's sophisticated adversaries tend to use several intrusion patterns to escape from modern IDS.

\begin{figure}[t]
\centering
\hspace*{-43pt}
\begin{tabular}{rl}
\includegraphics[width=0.6\linewidth,height=0.54\linewidth]{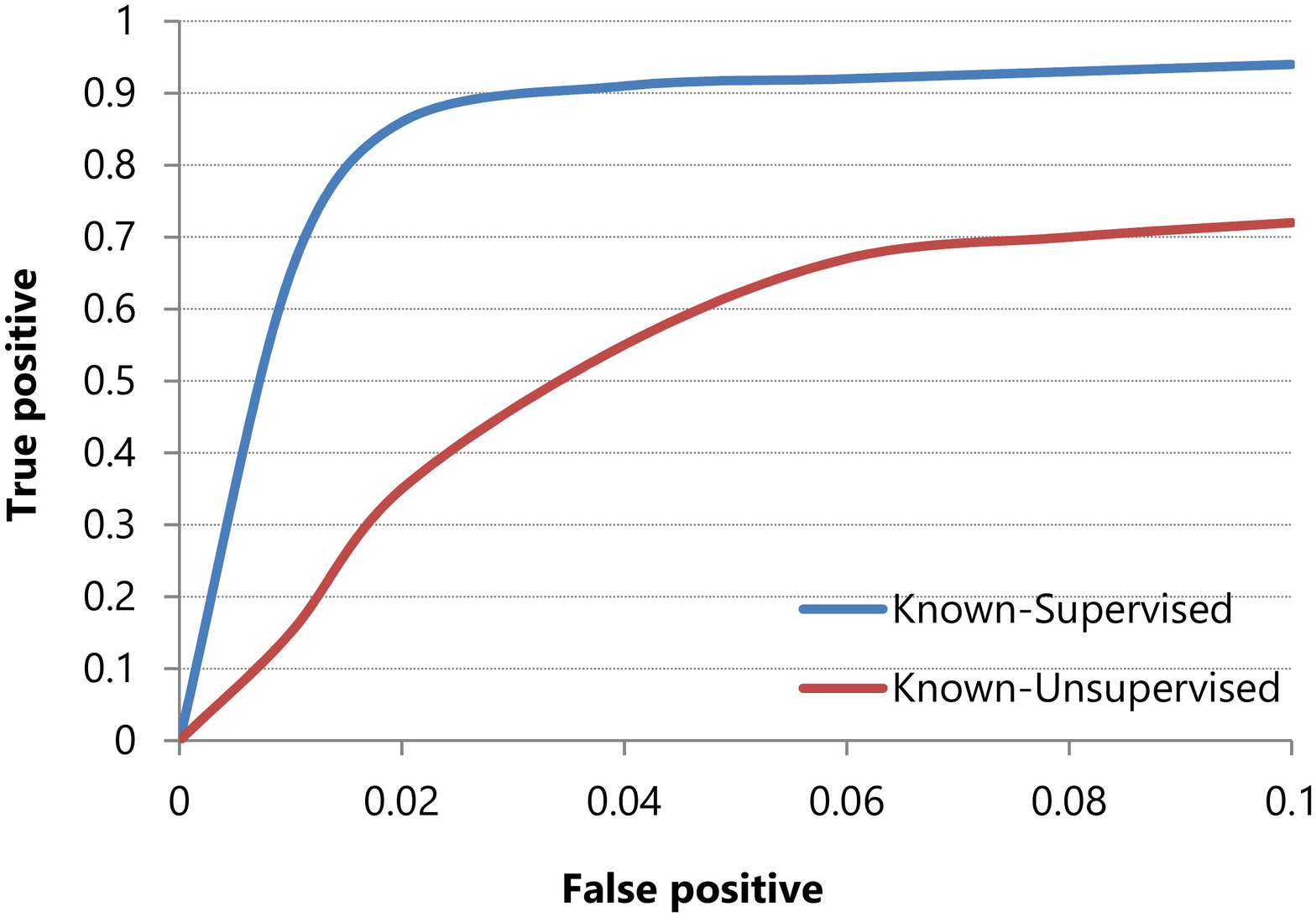} &  \hspace*{-30pt}\includegraphics[width=0.6\linewidth,height=0.55\linewidth]{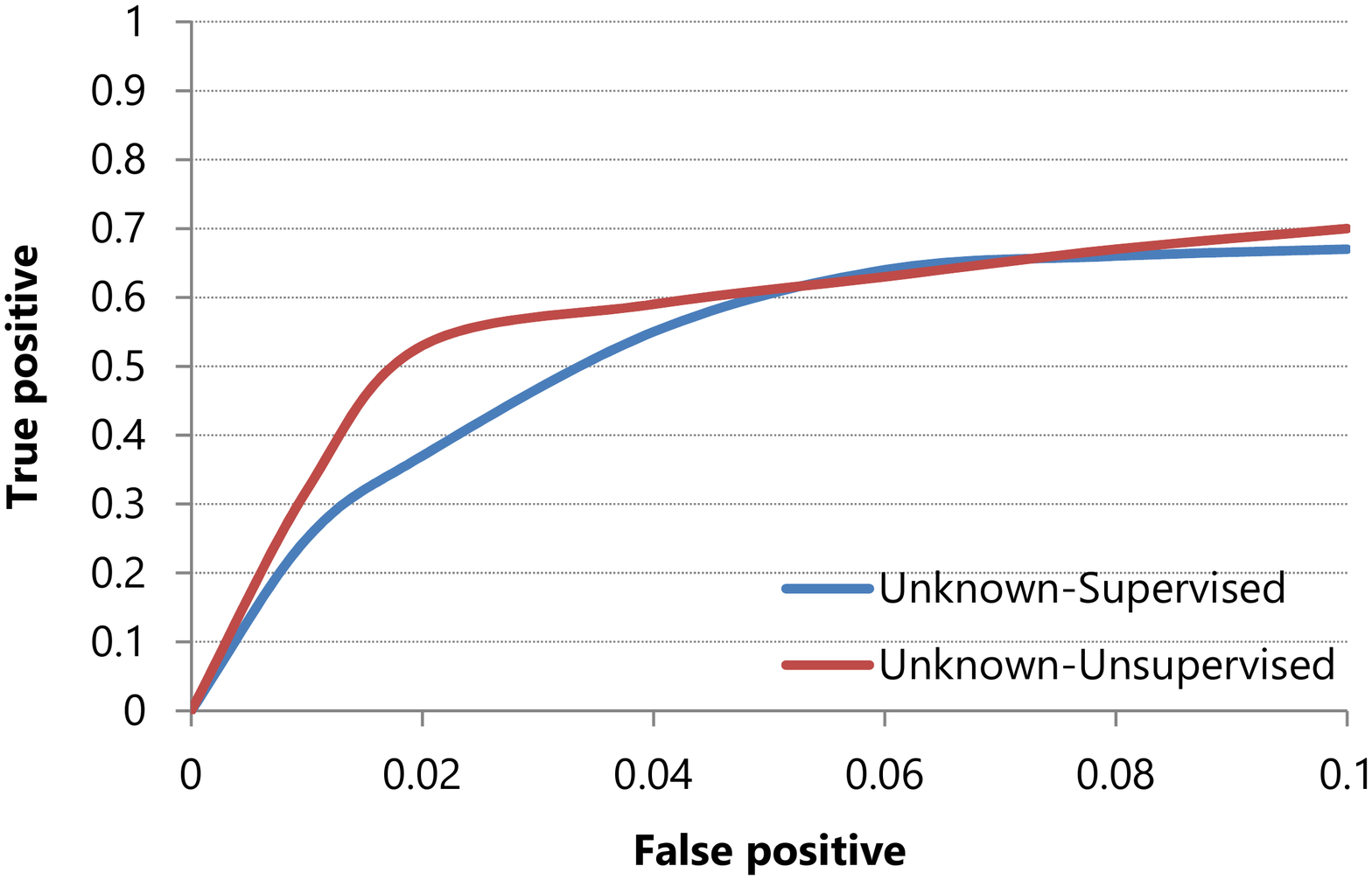}
\end{tabular}
\vspace*{-20pt}
\caption{Average of detection rates for methods evaluated in~\cite{Laskov:2005} in two scenarios: test data contains only known attacks (left) and test data contains unknown attacks (right).}
\label{fig:known}
\end{figure}

The results of~\cite{Laskov:2005} show that the supervised algorithms in general show better classification accuracy on the data with known attacks (the first scenario). Among these algorithms, the decision tree algorithm has achieved the best results (95\% true positive rate, 1\% false-positive rate). The next two best algorithms are the MLP and the SVM, followed by the $k$-nearest neighbor algorithm. However, if there are unseen attacks in the test data, then the detection rate of supervised methods decreases significantly. This is where the unsupervised techniques perform better as they do not show significant difference in accuracy for seen and unseen attacks. Figure~\ref*{fig:known} shows the average true/false positive rates of all methods evaluated in~\cite{Laskov:2005}. As the plots show, the supervised techniques generally perform better although unsupervised methods give more robust results in both scenarios.

Zanero and Savaresi~\cite{Zanero:2004:ULT:967900.967988} introduce a two-tier anomaly-based architecture for IDS in TCP/IP networks based on unsupervised learning: the first tier is an unsupervised clustering algorithm, which build small-size patterns from the network packets payload. In other words, TCP or UDP packet are assigned to two clusters representing normal and abnormal traffic. The second tier is an optimized traditional anomaly detection algorithm improved by the availability of data on the packet payload content. The motivation behind the work is that unsupervised learning methods are usually more powerful in generalization of attack patterns than supervised methods thus, there is a hope that such an architecture can resist polymorphic attacks more efficiently.

Lee and Solfo~\cite{Lee:1998:DMA:1267549.1267555} build a classifier to detect anomalies in networks using data mining techniques. They implement two general data mining algorithms that are essential in describing normal behavior of a program or user. They propose an agent-based architecture for intrusion detection systems, where the learning agents continuously compute and provide the updated detection models to the agents. They conduct experiments on Sendmail\footnote{Sendmail is an email routing API that supports several mail-transfer and delivery protocols such as SMTP used for email transport over the Internet.} system call data and network tcpdump data to demonstrate the effectiveness of their classification models in detecting anomalies. They finally argue that the most important challenge of using data mining approaches
in intrusion detection is that they require a large amount of audit data in order to compute the profile rule sets.

Sommer and Paxson \cite{Sommer10outsidethe} study the imbalance between the extensive amount of research on ML-based intrusion detection versus the lack of operational deployments of such systems. They identify challenges particular to network intrusion detection and provide a set of guidelines for fortifying future research on ML-based intrusion detection. More specifically, they argue that an anomaly-based IDS requires \emph{outlier detection}\footnote{Detecting deviations from known attacks.} while the classic application of ML is a classification problem that deals with finding similarities between activities. It is true that in some cases, an outlier detection problem can be modeled as a classification problem in which there are two classes: normal and abnormal. In machine learning, one needs to train a system with training patterns of \emph{all} classes while in anomaly detection one can only train on normal patterns. This means that anomaly detection is better for finding variations of known attacks, rather than previously unknown malicious activity. This is why ML methods have been applied to spam detection more effectively than to intrusion detection. 

\section{CI-based Techniques}
\label{sec:CI-Techniques}
In this section, we review several algorithms based on the four core techniques of computational intelligence: genetic algorithms (Section~\ref{sec:ga}), artificial neural networks (Section~\ref{sec:ann}), fuzzy logic (Section~\ref{sec:fl}), and artificial immune systems (Section~\ref{sec:ais}). 

\subsection{Genetic Algorithms (GA)} 
\label{sec:ga}
Genetic algorithms are aimed at finding optimal solutions to problems. Each potential solution to a problem is represented as a sequence of bits (\emph{genes}) called a \emph{genome} or \emph{chromosome}. A genetic algorithm begins with a set of genomes (population) and an evaluation function called \emph{fitness function} that measures the quality (goodness) of each genome. The algorithm uses two reproduction operators called \emph{crossover} and \emph{mutation} to create new descendants (solutions), which are then evaluated. Crossover determines how various properties of the parents in a population are inherited by the descendants. Mutation is the spontaneous alteration of a single gene.

Sinclair \etal~\cite{Sinclair:1999:AML:784590.784673} use genetic algorithms and decision trees to create rules for an intrusion detection expert system, which supports the analyst's job in differentiating anomalous network activity from normal network traffic. In this work, GA is used to evolve simple rules for network traffic. Each rule is represented by a genome and the initial population of genomes is a set of random rules. Each genome is comprised of 29 genes: 8 for source IP, 8 for destination IP, 6 for source port, 6 for destination port, and 1 for protocol. 

The fitness function is based on the actual performance of each rule on a pre-classified data set. An analyst marks a data set comprised of connections as either normal or abnormal. The system uses analyst-created training sets for rule development and analyst decision support. If a rule completely matches an abnormal connection, then it is rewarded a bonus and if it matches a normal connection it is penalized. Hence, the generations are biased toward rules that match intrusive connections only. Once the genetic algorithm reaches a certain number of generations, it stops and the best genomes (\ie rules) are selected. The generated rule set can be used as knowledge inside the IDS for judging whether the network connection and related behaviors are potential intrusions.

The traditional GA tends to converge to a single best solution called \emph{global maximum}. Since, the algorithm of~\cite{Sinclair:1999:AML:784590.784673} requires a group of best unique rules, a nature-inspired technique called \emph{niching} that attempts to create subpopulations which converge on local maxima. Details of various niching techniques are described in~\cite{542701}.

Li~\cite{wei:2004} describes a few disadvantages of the algorithm proposed in~\cite{Sinclair:1999:AML:784590.784673} and defines a new technique for defining IDS rules. They argue that in order to detect 
intrusive behaviors for a local network, network connections should be used to define normal and abnormal behaviors. An attack can sometimes be as simple as scanning for available ports in a server or a password-guessing scheme. But typically they are complex and are generated by automated tools. So, one needs to use temporal and spatial information of network connections to define IDS rules that can classify complex anomalous activities using an efficient genetic algorithm.

\subsection{Artificial Neural Networks (ANN)}
\label{sec:ann}
A neural network consists of a collection of processing units called \emph{neurons} that are highly interconnected according to a given topology. ANN have the ability to learning by example and generalize from limited, noisy, and incomplete data. They have been successfully employed in a broad spectrum of data-intensive applications.

Mukkamala \etal~\cite{Mukkamala:2002} describe approaches to intrusion  detection using neural networks and \emph{Support Vector Machines (SVM)}. Their goal is to discover patterns or features that describe user behavior to build classifiers for recognizing anomalies.
SVM are supervised learning machines that represent the training vector in high-dimensional feature space and label each vector by its class. SVM define an upper bound on the margin (separation) between different classes to minimize the generalization error, which is the amount of error in classification of unknown vectors. SVM classify data by determining a set of training data called support vectors that approximate a hyperplane in feature space.

Mukkamala \etal~\cite{Mukkamala:2002} use an SVM for non-linear classification\footnote{When the set of normal and abnormal points in a feature space cannot be divided with a line (or hyperplane), a non-linear function (hypersurface) is required to separate the two classes.} of feature vectors in an IDS. The SVM is trained with 7312 data points and test with 6980 test points from KDD\footnote{Knowledge Discovery and Data Mining (KDD) is a set of benchmark data created by MIT's Lincoln Labs and is considered a standard benchmark for intrusion detection. It contains raw  TCP/IP dump data for a local-area  network (LAN) that is under several attacks.}. Each point is located on a 41-dimensional space and the training is done using the \emph{radial basis function (RBF)}\footnote{Any function $\phi$ that satisfies the property $\phi(\mathbf{x}) = \phi(\|\mathbf{x}\|)$ is a \emph{radial function}, where $\|x\|$ is the distance of $x$ from an origin.}. The RBF is used to approximate the non-linear hyperplane that separates the normal and abnormal classes. Using this SVM, they reach an accuracy of 99.5\% in classification of test points. They also use three multi-layer feed-forward ANN to classify the same test points. The ANN are trained using the same 7312-point training set. The best result from experimenting the different ANN architectures is a detection rate of 99.25\%. The authors conclude that although their SVM IDS shows higher detection rates than their ANN, SVM can only be used for binary classification, which is a big limitation for IDS that require multiple classes.

\subsection{Fuzzy Logic}
\label{sec:fl}
Fuzzy logic is a method to computing based on \emph{degrees of truth} rather than the usual true or false Boolean logic on which the modern computers are based. With fuzzy spaces, fuzzy logic allows an object to belong to different classes at the same time. This makes fuzzy logic a great choice for intrusion detection because the security itself includes fuzziness and the boundary between the normal and anomaly is not well defined. Moreover, the intrusion detection problem involves many numeric attributes in collected data, and various derived statistical measures. Building models directly on numeric data usually causes high detection errors. An behavior that deviates only slightly from a model may not be detected or a small change in normal behavior may cause a false positive. With fuzzy logic, it is possible to model this small deviations to keep the false positive/negative rates small. 
Every fuzzy rule has the following general form,
\begin{center}
\texttt{IF \textsf{condition} THEN \textsf{conclusion} [\textsf{weight}]},
\end{center}
where \textsf{condition} is a fuzzy expression defined using fuzzy logic operators like fuzzy AND and fuzzy OR, \textsf{conclusion} is an atomic expression, and \textsf{weight} is a real number in $[0,1]$ that shows the confidence of the rule.

Gomez and Dasgupta~\cite{gomezevolving:2002} show that with fuzzy logic, the false alarm rate in determining intrusive activities can be reduced. They define a set of fuzzy rules to define the normal and abnormal behavior in a computer network, and a fuzzy inference engine to determine intrusions. They use a genetic algorithm to generate fuzzy classifiers, which is a set of fuzzy rules in the form defined above. Each fuzzy rule is represented by a genome and the GA is used to find the best genomes (fuzzy rules) to be added to the fuzzy classifier. The authors conducted experiments using the KDD evaluation data to classify 22 different types of attacks into 4 intrusion classes: denial of service (DoS), unauthorized access from a remote machine (R2L), unauthorized access to local superuser (root) privileges (U2R), and probing (PRB). The results show that their algorithm achieves an overall true positive rate of 98.95\% and a false positive rate of 7\%.

\subsection{Artificial Immune Systems (AIS)}
\label{sec:ais}
Natural immune systems consist of molecules, cells, and tissues that establish body's resistance to infections caused by pathogens like bacteria, viruses, and parasites. They distinguish pathogens from self cells and eliminate the pathogens. This provides a great source of inspiration for computer security systems, especially IDS. An artificial immune system is a computationally intelligent system based on behavior of the natural immune systems.

The first immune-inspired model applicable to various computer security problems was proposed by Hofmeyr and Forrest~\cite{Hofmeyr:1998:IDU:1298081.1298084}. Their model is specialized to detect intrusions in local area networks based on TCP/IP. They build a database containing normal sequences of system calls that act as the \emph{self} definition of the normal behavior of a program, and as the basis to detect anomalies\footnote{This work later inspired the work of~\cite{Lee:1998:DMA:1267549.1267555} described in Section~\ref{sec:AI-Techniques}.}. Each TCP connection is modeled by a triple, which encodes address of sender, address of receiver and port number of the receiver. Detectors are generated randomly through negative selection algorithm (NSA). In addition to NSA that results in a signal to stimulate or tolerate the immune response, they used a second signal (called co-stimulation) to confirm the anomaly that was detected through NS procedure. In this system, a human is required to generate this signal manually in order to reduce false alarms (autoimmunity) of the system.

Kim \etal~\cite{Kim:2007:ISA:1298636.1298643} provide an introduction and analysis of the key developments within the field of immune-inspired computer security as well as suggestions for future research. They summarize six immune features that are desirable for an effective IDS: distributed, multi-layered, self-organised, lightweight, diverse and disposable. They explain that the human immune system is distributed through immune networks and it generates unique antibody sets to provide the first four requirements. It is self-organized through gene library evolution, negative selection, and clonal. Finally, it is lightweight through approximate binding, memory cells, and gene expression to increase efficiency. 

Zamani \etal~\cite{Zamani:2009:DIM:1723200.1724175,DBLP:journals/corr/ZamaniMEP14} describe an artificial immune algorithm for intrusion detection in distributed systems based on danger theory, a immunological model based on the idea that the immune system does not recognize between self and non-self, but rather between events that cause damage. The authors propose a multi-agent environment that computationally emulates the behavior of natural immune systems is effective in reducing false positive rates. They show the effectiveness of their model in practice by performing a case study on the problem of detecting distributed denial-of-service attacks in wireless sensor networks.

Dasgupta~\cite{dasgupta99} proposes a multi-agent IDS based on AIS. He defines three types of agents: monitoring agents that roam around the network and monitor various parameters simultaneously at multiple levels (user to packet level), communicator agents that are used to play the role of signals between immune cells called lymphokines and decision/action agents to make decisions based on collected local warning signals. Roles of each type of agents is unique, though they may work in collaboration. This work unfortunately does not provide any experimental results making it difficult for the reader to compare the performance of the proposed system with other ML-based IDS. 

\section{Conclusion}
We reviewed several influential algorithms for intrusion detection based on various machine learning techniques. Characteristics of ML techniques makes it possible to design IDS that have high detection rates and low false positive rates while the system quickly adapts itself to changing malicious behaviors. We divided these algorithms into two types of ML-based schemes: Artificial Intelligence (AI) and Computational Intelligence (CI). Although these two categories of algorithms share many similarities, several features of CI-based techniques, such as adaptation, fault tolerance, high computational speed and error resilience in the face of noisy information, conform the requirement of building efficient intrusion detection systems. 

\bibliographystyle{plain}
\bibliography{refs}

\end{document}